# Nonlinear Processing with Linear Optics

Mustafa Yildirim[1,2,†], Niyazi Ulas Dinc[1,2,†], Ilker Oguz[1,2], Demetri Psaltis[2] and Christophe Moser[1]

[1] Laboratory of Applied Photonics Devices, Ecole Polytechnique Fédérale de Lausanne (EPFL), Switzerland

[2] Optics Laboratory, Ecole Polytechnique Fédérale de Lausanne (EPFL), Switzerland

† Equally contributing authors

## Abstract

Deep neural networks have achieved remarkable breakthroughs by leveraging multiple layers of data processing to extract hidden representations, albeit at the cost of large electronic computing power. To enhance energy efficiency and speed, the optical implementation of neural networks aims to harness the advantages of optical bandwidth and the energy efficiency of optical interconnections. In the absence of low-power optical nonlinearities, the challenge in the implementation of multilayer optical networks lies in realizing multiple optical layers without resorting to electronic components. In this study, we present a novel framework that uses multiple scattering that is capable of synthesizing programmable linear and nonlinear transformations concurrently at low optical power by leveraging the nonlinear relationship between the scattering potential, represented by data, and the scattered field. Theoretical and experimental investigations show that repeating the data by multiple scattering enables non-linear optical computing at low power continuous wave light. Moreover, we empirically found that scaling of this optical framework follows the power law as in state-of-the-art deep digital networks.

Optical computing has reemerged as an alternative to electronics for performing computations and handling information, particularly in the context of artificial intelligence applications. Optical neural networks (ONNs) hold promise in terms of speed and energy efficiency compared to traditional electronic computing[1]. However, the development of fully optical ONNs has proven to be a challenging task due to the need for incorporating both linear and nonlinear computations within the optical domain[2]. While several approaches have demonstrated efficient optical computing hardware for linear calculations[3-8], effectively integrating these capabilities with nonlinear computations remains a significant obstacle for a complete realization of ONNs. Researchers have explored nonlinear light-matter interactions in the context of reservoir computing[9-11], employing high intensity pulsed lasers for nonlinear data processing[12-15]. Additionally, the complex dynamics observed in multimode laser cavities interacting with external optical signals are employed for low-power nonlinear transformations in reservoir computing[16]. Platforms like integrated meshes of Mach Zehnder Interferometers[3], diffractive neural networks[6,17-18], micro-ring resonators[7,19], and free space linear systems[5,8,20,21] have facilitated linear calculations. However, for nonlinear computations, optoelectronic nonlinearity or electronic computation has been relied upon, resulting in limitations such as non-programable optoelectronic nonlinearity and high energy consumption. Therefore, there is a need to find a low power flexible solution to implement programable non-linear operations in the optical domain to fully harness the low power computing potential offered inherently by linear optics.

It has been previously shown that a purely linear transformation can be implemented with a stack of 2D diffractive layers[22]. The Ozcan group has applied this approach to ONNs employing additive manufacturing techniques[6,18]. These deep learning-enabled multi-layer diffractive processors enable computation by facilitating the propagation of free-space light through a sequence of structured passive scattering surfaces. This optical processing technique leverages the three-dimensional connectivity between nodes in consecutive layers, achieved via diffraction, thereby providing a path to scalability[23]. However, one limitation of this approach is that the nonlinearity is limited to the square law detection at the output which limits the realization of complex ONNs.

Another avenue that can be explored is the relationship between the scattering potential and the scattered light. While at low intensity levels, the propagation of light through a scattering medium exhibits linearity in terms of the relation between input and output light field, the output light can have a nonlinear dependence on the data encoded in the scattering potential. This form of nonlinearity is referred to as structural nonlinearity, and it has been investigated by Eliezer et al. using multiple scatterings within an integrating sphere[24].

In this paper, we present a programmable framework called nonlinear Processing with only Linear Optics (nPOLO) for the all-optical realization of neural networks using a low-power continuous wave laser and diffractive layers. The nPOLO framework enables simultaneous linear and nonlinear operations within the optical domain. In this way, nPOLO unifies multi-layer light modulation and structural nonlinearity such that the collective impact of data modulated layers on propagating light generates high order nonlinear transform of the data. Data is repetitively embedded into the modulation layers, combined with trainable parameters that enable the desired relationship (linear and nonlinear) between the data and the output field. Our results demonstrate that increasing the number of layers and data repetitions leads to the generation of higher-order nonlinearities, such as polynomial orders, which include cross-terms among the different elements of the input data. To illustrate the effectiveness of data repetition, we conducted a comparative analysis of the performance obtained between repeating the data in each modulation layer and presenting the data only once. Our results demonstrate that, when both systems have an equal number of degrees of freedom in terms of the displayed pixels in modulation layers, the data repetition approach consistently achieves higher accuracy scores and exhibits improved robustness against experimental imperfections and simulated noise. Overall, our findings showcase the ability of the nPOLO framework to synthesize a learnable both linear and non-linear data transform in a hybrid optical-digital neural network using only low power continuous wave light.

## nPOLO framework

The core of the nPOLO technique involves utilizing multiple data planes that are evenly spaced apart. The physical implementation of nPOLO includes a liquid crystal Spatial Light Modulator (SLM) and a mirror positioned opposite to it[25], allowing the simultaneous display of multiple modulation planes on a single SLM device. This configuration forms a multi-bounce single-pass cavity, where each plane serves as a reflecting surface that modulates the phase of light as it propagates from one plane to the next. Fig. 1 provides an unfolded representation of the multi-bounce nPOLO architecture. We can write the output field as a function of the modulation layers as follows:

$$E_{\text{out}} = HT_{LN}(\mathbf{x})\dots HT_{L2}(\mathbf{x})HT_{L1}(\mathbf{x})E_{il} \qquad \text{Eq. (1)}$$

Where $E_{il}$ is the illumination beam, $T_{Ln}$ is the transmittance of the *n*th modulation layer, where the modulation layer number goes up to $N$. $H$ is the diffraction operator that is the same for all the layers as they are equally spaced apart. Without loss of generality, let us assume that we have an optical system that has an input aperture and output aperture that are sampled by total number of $K$ pixels. Then the input and output electric fields become $K \times 1$ vectors. We can write the incident field as a vector of ones, assuming the incident optical field is a plane wave propagating along the optical axis. Assuming the same number of pixels in the modulation layer as in the input/output apertures, $T_{Ln}$ becomes a diagonal matrix with size $K \times K$. $H$ is a $K \times K$ Toeplitz matrix to represent diffraction for an arbitrary distance[23]. If one inserts directly the data to $T_{Ln}$, there is no trainable parameter to tune the transform. Thus, instead of the direct use of the data in $T_{Ln}$, we feed a linearly transformed version of it in the form of $t_j^{(n)} = s_j^{(n)} x_j^{(n)} + b_j^{(n)}$ to introduce trainable parameters where $s_j^{(n)}$ is a scaling parameter, $x_j^{(n)}$ is an element of data vector and $b_j^{(n)}$ is a bias parameter where $j$ refers to indexing of the vectors and $n$ is the layer index. When we expand Eq. (1) for the general case of complex modulation with these parameters, we get:

$$o_i = \sum_{j=1}^{K} h_{ij}\left(s_j^{(N)}x_j + b_j^{(N)}\right)\left(...\sum_{j=1}^{K} h_{ij}\left(s_j^{(2)}x_j + b_j^{(2)}\right)\left(\sum_{j=1}^{K} h_{ij}\left(s_j^{(1)}x_j + b_j^{(1)}\right)\right)...\right) = \sum_{m=0}^{N} \alpha_{i,m}(\mathbf{s}, \mathbf{b})\mathbf{x}^m \quad \text{Eq. (2)}$$

Where $h_{ij}$ is the element of the diffraction matrix. As seen by the left-hand side of Eq. (2), multiplication with **x** at each layer gives rise to polynomial orders. Therefore, we can write the overall transform in a generic polynomial form up to order $N$ where coefficients ($\alpha_{i,m}$) are functions of **s** and **b**, the right-hand side of Eq. (2).

Now, we look at how the input is transformed by the nonlinear activation function applied to the linear transformation of **x** for a perceptron in a digital neural network:

$$o_i = g\left(\sum_{j=1}^{K} w_{ij}x_j + b_i\right) = \sum_m \beta_{i,m}(\mathbf{w}, \mathbf{b})\mathbf{x}^m \quad \text{Eq. (3)}$$

In Eq. (3), $w_{ij}$ and $b_i$ are trainable weight and bias parameters of the perceptron, and $g$ is an algorithmic nonlinear function of choice (sigmoid, ReLU, etc.). The nonlinear activation function can be expressed as a polynomial expansion as well, where the coefficients ($\beta_{i,m}$) depend on **w** and **b**. In a Multi-Layer Perceptron (MLP), the output is a cascade of such polynomials which is also a polynomial. Therefore, the functional form of Eq. (3) applies to an MLP as well. Eq. (2) and Eq. (3) have a similar polynomial form except with different coefficients. Ideally, if the nPOLO and the MLP are both trained perfectly to implement the same function, then the coefficients of the two polynomial expansions should converge. In the Supplementary Material (Section 1), we present fully developed output vector **o** in terms of explicit polynomial orders.

Trainable parameters, in the form of scaling and bias, are applied to each pixel value of the data presented on the modulation layers. These parameters are trained digitally via a computer model (see Methods). Once the desired nonlinear transformation is achieved on the computer, the parameters are applied to the multiple layers (adjacent planes on the SLM) in the experimental setup, resulting in an intensity pattern that is recorded by a camera. Subsequently, a compact representation of the recorded camera pattern is obtained through average pooling, resulting in a 2D matrix of values such as a 4-by-4 or an 8-by-8 grid. This compact representation is then fed into a digital linear classifier, which processes the data via a single fully connected linear layer, thereby producing the final classification results.

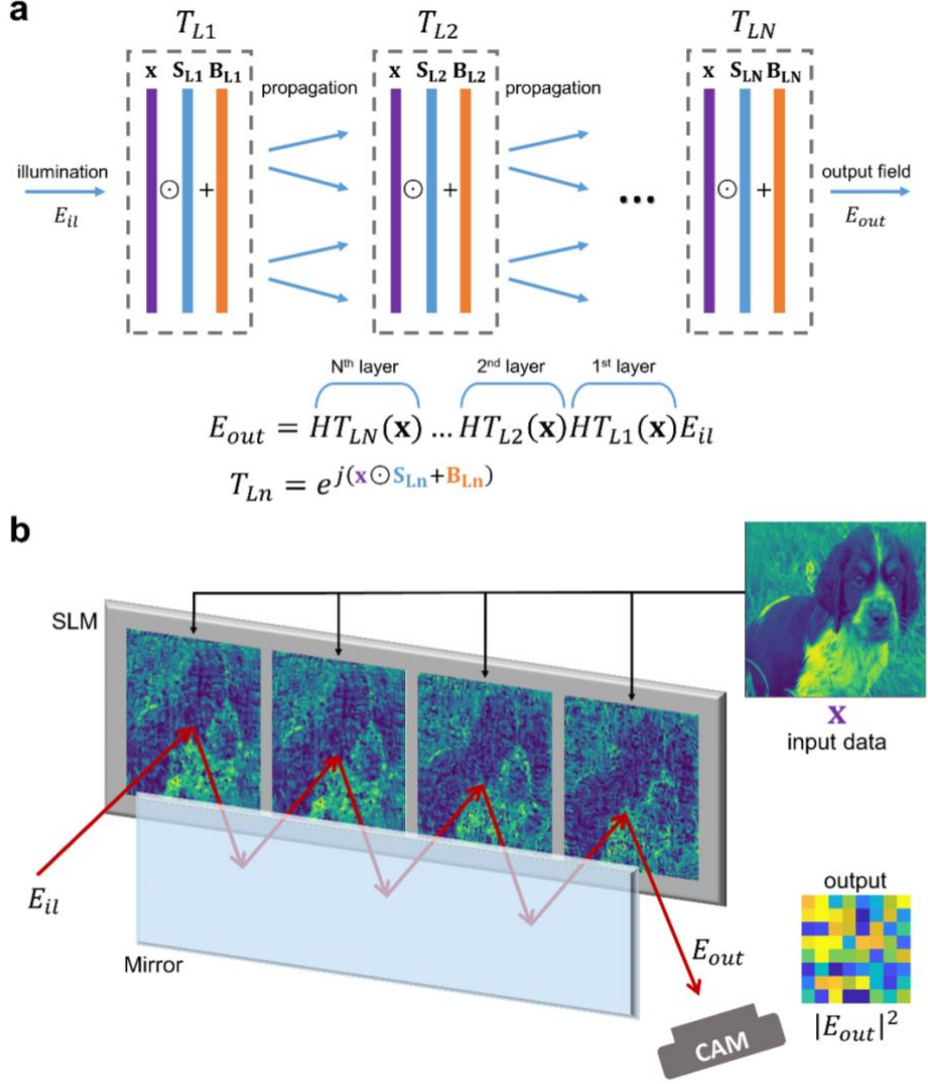

*Figure 1. The framework of nPOLO* a) The computation scheme is depicted, showcasing multiple modulation layers ($T_{Ln}$) within the framework. The data (**x**) is presented on these layers, accompanied by trainable scaling ($S_{Ln}$) and bias ($B_{Ln}$) parameters to optimize the transformation. Propagation is represented by H (Toeplitz matrix) b) The physical implementation nPOLO, featuring a single pass multi-bounce cavity configuration. This implementation consists of a Spatial Light Modulator (SLM) and a mirror, positioned in a way that enables the realization of consecutive layers on the SLM side by side. The propagation distance is determined by the reflection from the mirror, allowing the light to propagate between the layers. Output light is captured with a camera.

## Results

By increasing the number of layers i.e. adjacent planes on the SLM, one can assess the impact of the polynomial orders resulting from structural nonlinearity. However, the increase in the number of layers also leads to an increase in the system's degrees of freedom and space-bandwidth product. Therefore, we devised an alternative comparison experiment to mitigate these effects. In our experiments, we maintained the same number of layers and pixels but modified the data allocation. Specifically, we initially incorporated the data only in the first layer, while the subsequent layers consisted exclusively of trainable bias parameters without any data or scaling parameters. To evaluate the performance of the nPOLO framework, we conducted experiments using the Imagenette, Fashion MNIST and Digit MNIST datasets[26-28]. The obtained numerical and experimental results are provided in Fig. 2.

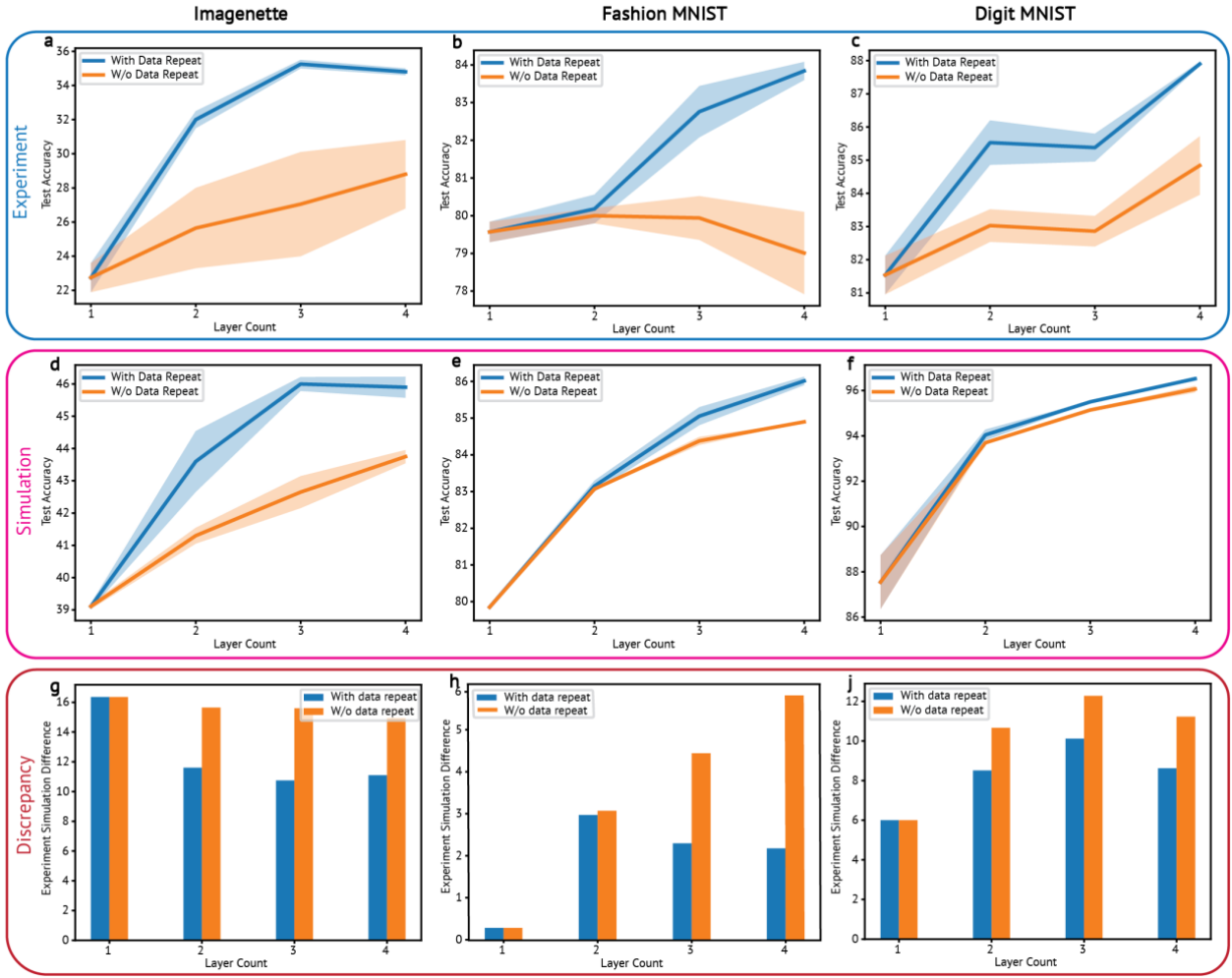

*Figure 2 showcases the classification accuracy results obtained for the Imagenette (a, d, g), Fashion MNIST (b, e, h) and Digit MNIST (c, f, j) datasets, comparing two different schemes: one "with data repeat" and one "without data repeat".* The layer count (N) varies from one to four for both schemes. Each configuration is trained independently, resulting in layer masks that are applied to the Spatial Light Modulator (SLM) as phase masks. a, b, c) The experimentally obtained test accuracies for all datasets are displayed, representing the performance with and without structural nonlinearity. d, e, f) Test accuracies of corresponding simulations are plotted for both schemes likewise in experiments. The mean and standard deviation values are obtained by testing three times the models trained from scratch. g, h, j) The mean accuracy difference between experimental and simulated results are shown as bar plots for varying layer number with and without data repeat.

We use 300-by-300 pixels on SLM for each modulation layer whereas the original Fashion and Digit MNIST dataset samples contain 28-by-28-pixel images and Imagenette dataset samples contain 320-by-320-pixel images. For the Fashion and Digit MNIST datasets, we used 4-by-4 superpixels on the SLM, yielding 75-by-75 grid for assigning trainable parameters, which yields 11250 parameters (scaling and bias) per layer. We accordingly linearly up-sampled the images of those datasets to 75-by-75. For the Imagenette dataset, we linearly down-sampled the images to 300-by-300 and used all the pixels for assigning trainable parameters, yielding 180000 parameters (scaling and bias) per layer.

For clarity, we provide examples of the displayed masks in Fig. 3 "with data repeat" and "without data repeat" configurations using an example from the Fashion MNIST and Imagenette datasets with four modulation layers. Also see the Supplementary Material (Section 2) for the comparative depiction of parameter allocation. The trainable parameters were optimized by computer simulation, wherein the physical light propagation was modeled using the Beam Propagation Method (BPM).

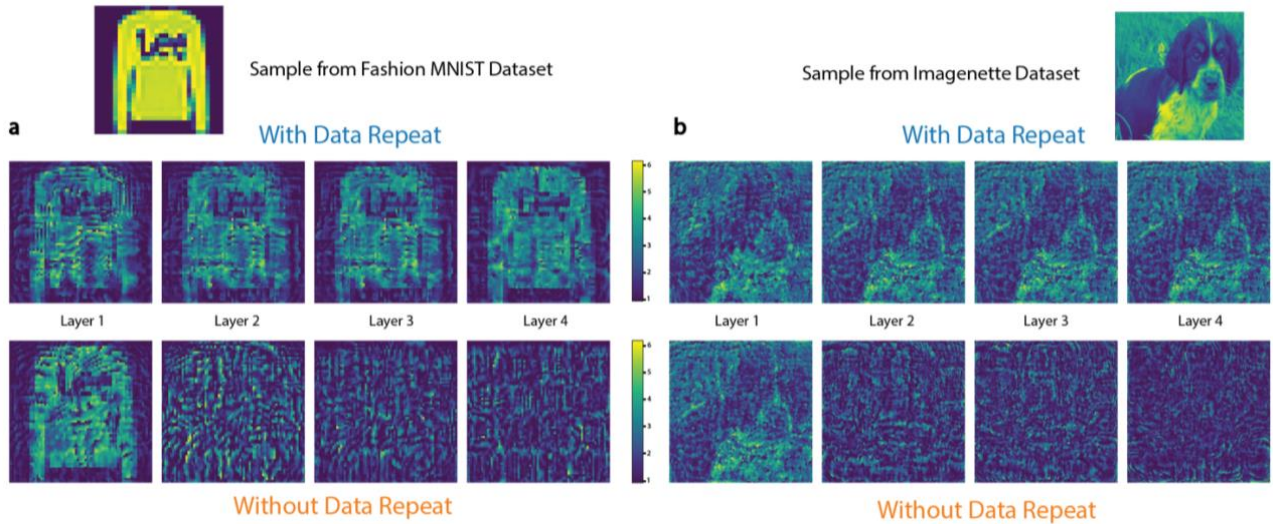

*Figure 3. Samples of trained layer masks to be displayed on SLM for Fashion MNIST and Imagenette.* *a) Presents an example sample from the Fashion MNIST dataset alongside its corresponding layer masks, combined with trainable parameters, according to both schemas. The color code utilized represents the phase modulation ranging from 0 to 2π. Similarly, panel (b) showcases an example sample from the Imagenette dataset, and its corresponding layer masks combined with trainable parameters for both schemas.*

Since BPM consists of differentiable calculation steps, the error can be backpropagated to the trainable parameters, and they are optimized using stochastic gradient descent. In Fig. 4, we present the training scheme used, in which the digital model of the optical system and the digital classifier were co-trained for the classification tasks. By following this co-training approach, we obtained scaling and bias masks for different layer configurations, ranging from layer $N = 1$ to layer $N = 4$, both with and without data repetition. It is important to note that in the case of a single layer ($N = 1$), both data repetition options are equivalent, as we only had a single layer available to introduce the data.

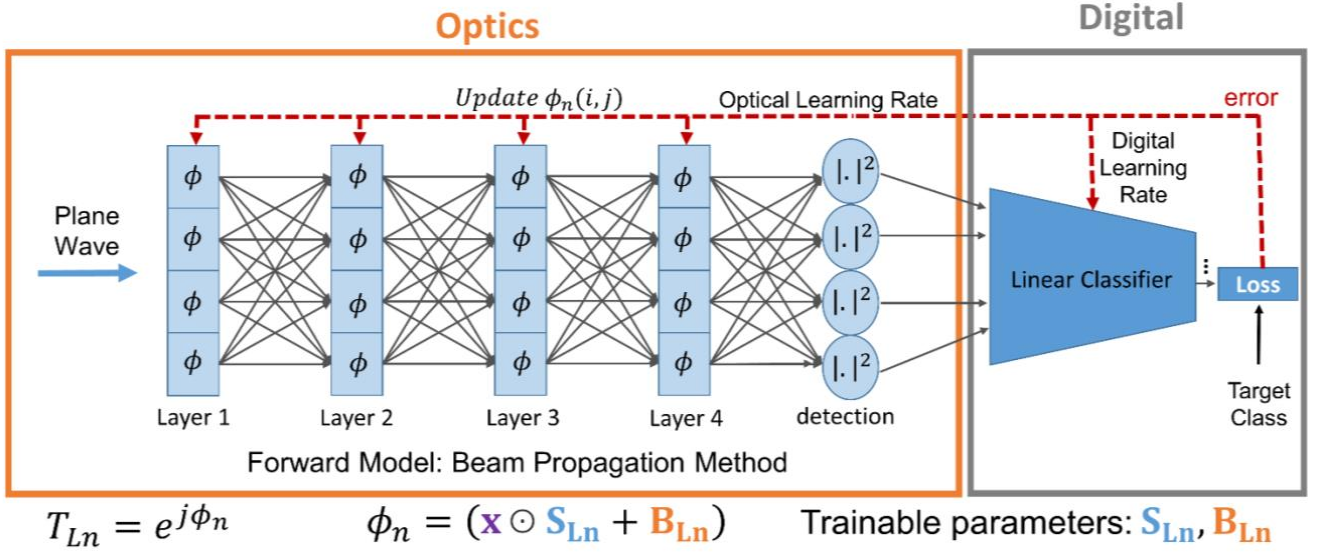

*Figure 4. Co-training of optical trainable parameters and digital trainable parameters.* Simulation model of four-layer system is built by beam propagation method, where layers are composed of trainable parameters ($S_{Ln}$ and $B_{Ln}$). Training of optical layers and linear classifier is performed simultaneously using separate learning rates.

Our experimental findings in Fig. 2a-c consistently demonstrated that when data was repeated across multiple layers, we achieved higher classification accuracy compared to configurations without data repetition. Moreover, increasing the number of layers also contributed to improved accuracy. We also observed that when the number of layers was held constant, eliminating data repetition led to a reduction in accuracy. These results highlight the contribution of higher order optical nonlinearities generated via data repetition. We observed a similar trend in our simulations while calculating the trainable parameters (see Fig. 2d-f). Both experimental and simulated results validate the contribution of the nPOLO framework. However, the experimental accuracies were lower than the simulated ones. Fig. 2g-j illustrate the accuracy difference between simulations and experiments in all three datasets. Interestingly, the decrease in accuracy is less pronounced in cases involving data repetition. We attribute this discrepancy to imperfections between the simulated model and the physical implementation, primarily the non-ideal phase response and flickering of the SLM. For example, we were able to reduce the discrepancy between the optical experiment and the digital simulation to 2% by upgrading the SLM for the Imagenette database with data repeat configuration (see Supplementary Material Section 3). Such engineering aspects of nPOLO framework will be further explored in a follow-up study. We also trained a simple convolutional neural network to assess how nPOLO compared to a fully digital counterpart for these tasks. and obtained comparable performance (see Supplementary Material Section 4).

## Scaling study

The performance is influenced by three factors: the number of trainable parameters, the dataset size, and the available compute budget for training[29]. While the dataset size remains constant in our study, the model size is subject to variation within the constraints of a limited computational budget. Specifically, we explore the impact of changing the model size on the performance of nPOLO, determined by the width, $W^2$, (the number of SLM pixels in the illuminated 2D patch) of its diffractive layers and the depth, $N$, denoting the number of diffractive layers. Our empirical investigation involved varying $W$ and $N$ for classification tasks of the Imagenette and Fashion MNIST datasets (see Methods). The findings of this study are summarized in Fig. 5. The parameter count of our model is $2NW^2$, where a factor of 2 is included due to the two trainable parameters per pixel. Initially, we observe a continuous upward trend in performance as the parameter count increases, regardless of $N$ and $W$. In other words, configurations with different depths perform similarly for the same parameter count, up to a few hundred thousand, as depicted in Fig. 5a and 5c. However, further increase in the parameter count leads to performance saturation. This phenomenon is also observed in conventional digital neural networks. Deeper implementations of nPOLO ($N > 1$) extend the onset of this saturation point. Recent studies at Open AI have revealed that the scaling of deep multilayer perceptron networks adheres to a power-law relationship between the test loss and the size of the model[29, 30]. The test loss of the simulated nPOLO is plotted in Fig. 5b and 5d against the number of parameters counts, revealing a similar power-law scaling.

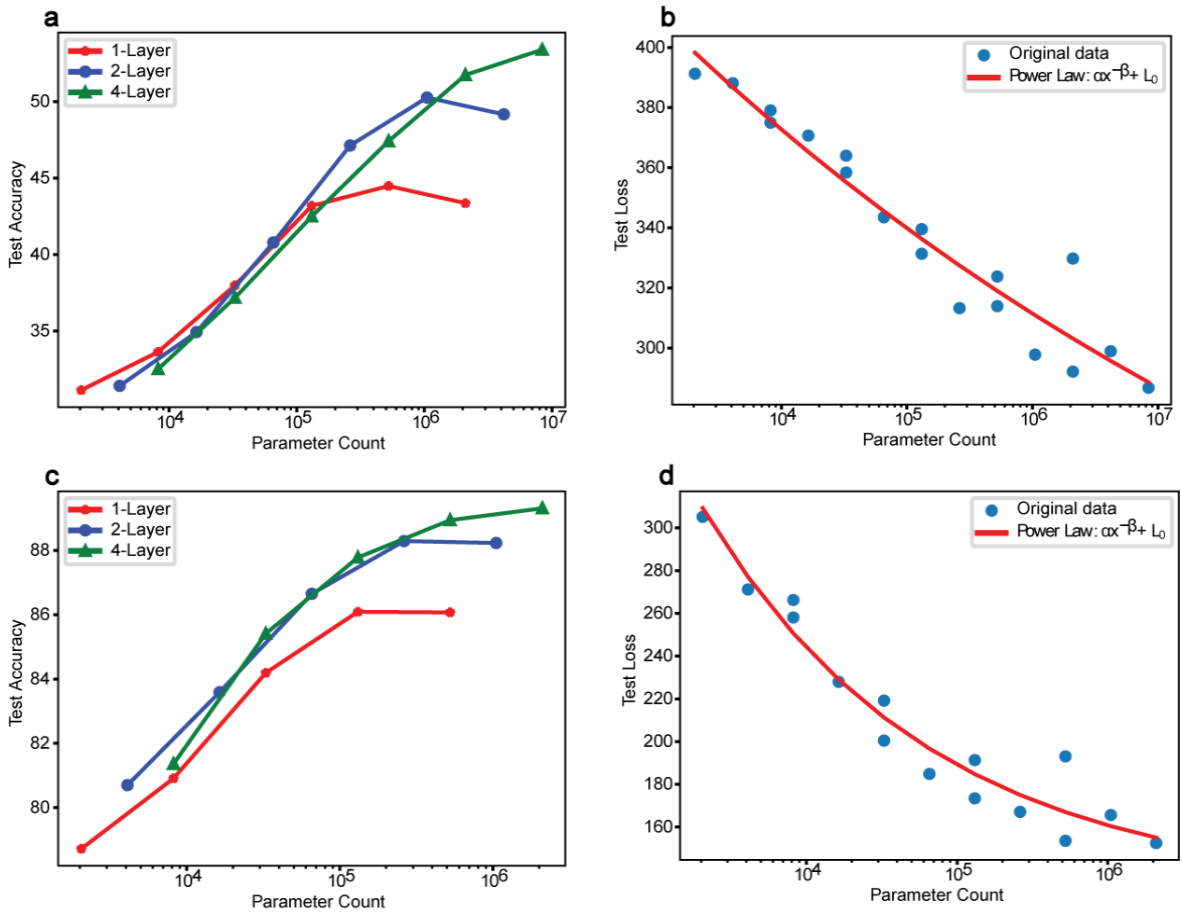

*Figure 5 illustrates the scaling dynamics of the nPOLO with increasing parameter count, showcasing test accuracies and losses for both the Imagenette (a, b) and Fashion MNIST (c, d) datasets across various*

***parameter configurations****. The manipulation of layer width and depth enables the realization of different parameter counts. Specifically, plots (a) and (c) provide a detailed breakdown of test accuracies for different layer counts (1, 2, and 4) individually. On the other hand, plots (b) and (d) present all test results in relation to parameter count, irrespective of the specific layer number.*

# Discussion

We introduced trainable scaling and bias parameters for the pixel values of the data displayed on the SLM to synthesize a programmable computation. We observed that the contribution of structural nonlinearity became more pronounced when dealing with more challenging classification tasks. The impact of using repeated data versus not using repeated data was more significant for Imagenette compared to Digit MNIST. This disparity arises because Digit MNIST represents an easier task, and the structural nonlinearity becomes redundant in the presence of the square law detection nonlinearity.

The scaling study we performed revealed that as the model size of nPOLO increases the performance improves and the optimal performance relies on a balanced scaling of width and depth of the model. Remarkably, similar properties are also found in digital deep neural networks[30,31]. Future studies could explore alternative approaches of employing trainable parameters or implementing different functional forms, such as using convolutional kernels or eigenmodes of the optical system as trainable parameters, similar to investigations conducted for fiber-based optical learning machines[32]. During the final stage of this work[33], we became aware of an independent and different approach to perform passive optical non-linearity, exploiting reflections inside a disordered cavity[34]. More recently another paper appeared investigating, through digital only simulations, structural nonlinearity for the implementation of neural networks[35].

An interesting observation is the increased robustness of data repetition across multiple layers against experimental imperfections. We initially noticed this phenomenon during experiments conducted on different datasets and layer configurations. To further investigate this, we introduced phase noise and misalignment in BPM simulations to emulate experimental imperfections while keeping the trained masks fixed (see Supplementary Material Section 3). Gradually increasing the simulated noise level or misalignment degree, we observed that the configurations with data repetition exhibited greater robustness. The configuration without data repetition experienced a more rapid drop in accuracy, consistent with the experimental results. This finding strengthens the argument for the noise robustness of the data repetition scheme, as we have empirical data from both experiments and simulations. One possible explanation for this phenomenon is that by introducing the data multiple times, the model learns multiple paths from the input data to the output plane during training, resulting in not only higher polynomial orders but also cross-terms that couple with different optical paths and reach the detector plane. The existence of multiple routes for highlighting useful features in the output plane may make the data repetition scheme less susceptible to noise.

Our scaling study demonstrated a power-law scaling trend akin to the observations in OpenAI's deep neural network studies[29, 30]. While exploring deeper models with more than 4 layers for improved performance, we found that increased depth did not yield substantial enhancements. We hit the dataset bottleneck as we observe overfitting for large number of parameters that deeper models employ. The increased deviations towards high parameter count (Fig. 5) are also because of this reason. This aligns with the findings reported in references[29, 30] that empirical performance exhibits a power-law relationship with three factors of model size, dataset, and compute budget individually when not constraint by the other two. In practical terms, pursuing deeper models might offer diminishing returns unless accompanied by adjustments in dataset size and employing greater computational resources. We note that the highest parameter count in the scaling study yields ≈4.2 million pixels for Imagenette and ≈1 million for Fashion and Digit MNIST datasets, which can be employed using commercially available devices. Other possibilities to further scale the number of parameters include employing multiple SLMs or other structured media such as computer-generated holograms[36] or volume holograms[37, 38] instead of the flat mirror to accommodate additional parameters to boost the performance.

Overall, the nPOLO framework presents a novel approach for generating optical nonlinearity using low-power optical devices, eliminating the need for electronic components to achieve higher orders of nonlinearity. Networks implemented with structural nonlinearities are not multilayer perceptrons but they can be trained to reach comparable performance and they have similar scaling laws. Furthermore, we discovered that the introduction of data repetition to generate polynomial nonlinearities enhances robustness against noise. Note that this framework is applicable to the cascade of any optical linear system such as integrated waveguide Mach-Zehnder interferometers[3]. These characteristics make nPOLO a promising platform for realizing optical neural networks.

# Methods

## Digital training

Optimization methods are already demonstrated to reconstruct 3D phase objects from experimental recordings of 2D projections[39,40]. In ref[39], the forward model in the optimization is the beam propagation method (BPM). The iterative error reduction scheme and the multilayer structure of the BPM resembles a multilayer neural network. Therefore, this method is referred to as Learning Tomography. We showed that instead of imaging an object, we can reconstruct the 3D structure that performs the desired task as defined by its input-output functionality[36]. To establish the target functionality, the 3D phase modulation, either through a continuous medium or multiple planes, and the scattered field caused by the phase modulation must be accurately modeled. Unlike conventional reconstruction algorithms that rely on first-order approximations, LT incorporates higher-order scattering effects by employing BPM. The LT algorithm involves an iterative reconstruction process using the forward model, along with the constraints arising from experimental considerations such as the pixel pitch of the SLM. In this study, we adapted this approach presented in the ref[36], where additional details can be found, to generate scaling and bias parameters for demonstrated classification tasks. The output intensity pattern of the forward model is average pooled to yield a 4-by-4 matrix for each sample of Fashion MNIST dataset and an 8-by-8 matrix for each sample of the Imagenette dataset. These matrices are flattened to act as an input layer of a digital classifier that has 10 output neurons for each class of the datasets without any hidden layer and nonlinear activation function. The trainable parameters employed in the BPM model and digital classifier are co-trained by a continuous error backpropagation where different learning rates are assigned to digital weights (i.e., $10^{-4}$) and optical scaling and bias parameters (i.e., $10^{-3}$) using categorical cross entropy as the loss function. We used batch learning with a batch size of 20 and a random shuffle in every batch. PyTorch libraries are used for the whole training process.

For the scaling study, we followed an approach akin to classical neural networks by augmenting the number of trainable parameters in nPOLO. The parameter count per plane in nPOLO is W², where W signifies the plane's width. We conducted a sweep of W values from 32 to 1024 (up to 512 for Fashion MNIST) for bounce numbers of 1, 2, and 4. The test accuracies over 50 epochs for the Imagenette and Fashion MNIST are presented in Fig. 5. During scale-up studies we encountered overfitting and slight data augmentation (random flip and rotation) is introduced to reduce this.

## Experimental setup

A photograph showcasing our optical setup is presented in Supplementary Material Section 3. In our experiments, we employed a continuous wave Solstis M2 laser operating at a wavelength of $\lambda = 850 \, nm$. The mirror we used has a width of 11 mm, providing ample space for the four reflections. To deliver the beam to the SLM, we implemented 4f imaging to relay the beam reflected from a digital micromirror device (DMD). The use of a DMD offered the advantage of flexible beam sizing. Specifically, we configured the beam shape as a square with a side length of 2.4 mm, corresponding to a 300-by-300-pixel area on the SLM. It is worth noting that the SLM utilized in our setup has an $\Lambda = 8 \, \mu m$ pixel pitch. To construct a multi-bounce cavity, we utilized a Holoeye Pluto Spatial Light Modulator (SLM) and positioned the mirror at a distance of $d = 15.2 \, mm$ from the SLM screen. This distance is set so that diffraction from one corner pixel of a layer can reach to the opposite corner of the subsequent layer ($d \times \lambda/\Lambda \geq 2.4 \, mm$). The distance between SLM and mirror enables all to all pixel connectivity. The zero-order reflection from the SLM is used, which has 69% percent reflectivity. This configuration allowed the input beam to undergo four reflections from the mirror. Therefore, the total transmission due to the SLM efficiency after four bounces is 23%. After the fourth bounce, the beam on the SLM is magnified by a factor of 1.2, and the resulting output intensity is detected by a CMOS camera with a pixel pitch of 3.45 µm. The corresponding beam in the camera occupies an area of 834 by 834 pixels. During image acquisition, we applied average pooling to resize the obtained images to either 4-by-4 or 8-by-8 dimensions. For the Imagenette dataset, we used the whole training and test samples as originally prepared. For the Digit and Fashion MNIST datasets, we used the whole training set (60000 samples) for the simulations however we used the first 10000 samples for the re-training of digital weights after the experiments and we used the first 2500 samples of the test set for blind testing of the experimental results.

## Data availability

The data that support the plots within this article and other findings of this study are available from the corresponding author upon reasonable request.

## Code availability

All relevant code is available from the corresponding author upon reasonable request.

## Author contributions

M.Y., N.U.D., D.P., and C.M. conceived and initiated the project. N.U.D. provided the first version of numerical training module. M.Y. and N.U.D. modified the training scheme according to the needs of this project. M.Y., N.U.D., and I.O. worked on the experimental setup and used datasets. All the authors contributed to the guidance of the experiments and the discussion of the results. M.Y. and N.U.D. prepared the first draft of the manuscript. All the authors revised the manuscript. D.P. and C.M. supervised the project.

## Competing interests

The authors declare no competing interests.